# Security Rating Metrics for Distributed Wireless Systems

Volodymyr Buriachok[0000-0002-4055-1494], Volodymyr Sokolov[0000-0002-9349-7946], and Pavlo Skladannyi[0000-0002-7775-6039]

Borys Grinchenko Kyiv University, Kyiv, Ukraine
{v.buriachok,v.sokolov,p.skladannyi}@kubg.edu.ua

**Abstract.** The paper examines quantitative assessment of wireless distribution system security, as well as an assessment of risks from attacks and security violations. Furthermore, it describes typical security breach and formal attack models and five methods for assessing security. The proposed normalized method for assessing the degree of security assurance operates with at least three characteristics, which allows comparatively analyze heterogeneous information systems. The improved calculating formulas have been proposed for two security assessment methods, and the elements of functional-cost analysis have been applied to calculate the degree of security. To check the results of the analysis, the coefficient of concordance was calculated, which gives opportunity to determine the quality of expert assessment. The simultaneous use of several models to describe attacks and the effectiveness of countering them allows us to create a comprehensive approach to countering modern security threats to information networks at the commercial enterprises and critical infrastructure facilities.

## 1      Introduction

Several threats are presently affecting wireless systems: natural, man-made, human intentional and human inadvertent. Natural (cosmic radiation, ionization of the ionosphere) and man-made (radiation of radio equipment) are very similar in action: they cause interference in communication channels. Intentional threats become more widespread and appear as a form of security breaches: the introduction of malicious code into the system. Human inadvertent threats can be considered as force majeure [1, 2].

The reminder of the paper is organized as follows. Section 2 "Review of the Literature" contains the analysis of the latest scientific work in this area. Sections 3 "Problem Statement," 4 "Methods for Assessing the Threats," and 5 "Approaches to the Threats Assessment" reveal problems, well-known approaches to solving the problem of evaluating the effectiveness of information systems protection. In sections 6 "Zombie" Model of Security Breaches Consideration" and 7 "Formal Attack Model" are presented formal models of security breaches and attacks. In section 8 "Methods for Threats Assessing" considered the existing and charming own method of threats. Section 9 "Functional-Cost Analysis" is an example of an audit of the cost of a security system implementing. The paper ends with section 10 "Conclusion and Future Work."

## 2  Review of the Literature

At first glance, it seems that the problem of protection against security breaches can be solved by protecting the information, transmitted by the network, itself. But such a threat is due to the use of the computer facilities, directly involved into in data transmission, in equipment, for instance, multiplexers and demultiplexers, switches, routers, amplifiers, regenerators, control devices, etc. Thus, we are talking not only about the integrity of information, but also about the capacity of the system as a whole [1].

The system consists of hardware, software, information resources and organizational structure. Each of these elements can be considered separately as a subsystem of the general system and apply the same principles as for the system as a whole [3].

Theoretical and practical studies indicate that the determination of exact quantitative estimates of possible damage is very difficult or impossible at all. Due to this, the approximate estimates obtained during the operation of the wireless system, together with expert assessments have become widespread [4].

## 3  Problem Statement

Through the results of security breaches (cyber attacks and viruses) lead to a deterioration of the wireless system infrastructure, they can be considered similar to obstacles. Conversely, obstacles can be considered as the effects of viruses.

Simultaneously with the definition of security indicators, risk assessment should be considered. Only a combination of these two indicators provides a complete picture of the state of wireless system being studied.

The purpose of this study is to develop the methods for determining the level of security, refining them to obtain a methodology of comparing several systems among themselves, as well as improving the methods for verifying the reliability of expert assessments. Let's suppose that the attacker uses a known security breach model—a zombie model—to gain access to an object of information activity. Defining the system's security level can be used in conjunction with the tree method attack for timely response to changes in the system configuration, the emergence of new types of attacks and changes in organization security policy.

## 4  Methods for Assessing the Threats

*"Zombie" model* of security breaches consideration and *formal attack model* allow us to simulate a system and an attack on it. The most appropriate methods for assessing the threats from internal and external threats may be the following:

- *Denial of service probability*.
- *Expected vulnerability damage* from the $i^{th}$ threat.
- *Set of values for defining security requirements*.
- *Assessing the threats and losses*.
- *Degree of security procuring*.

Through empirical analysis of the above methods for assessing the threats, it has been found that none of its meets the requirements for security of information objects. In our view, this problem can be solved by means of the normalization of quantitative and qualitative indices of threats to information objects and, if necessary, used for "weakly structured" indicators of expert evaluation data. To this end, we have proposed a *comprehensive method* of the information objects security.

## 5 Approaches to the Threats Assessment

The subjective process of obtaining the probability of the threat can be divided into three stages:

- Preparatory (the object of research is formed: the set of events and the initial analysis of the properties of this set; one is selected for methods of obtaining subjective probability; the preparation of an expert or a group of experts is conducted).
- Derivation of the assessments (using the chosen method; obtaining results in a numerical form, possibly controversial).
- Analysis of the obtained assessments (researching the results of the survey; clarification of the experts' answers).

Sometimes the third stage is not carried out if the method itself uses the axioms of probable distribution, which is close to expert estimates itself. Conversely, the stage becomes especially important if results are obtained from expert groups [5].

It is also possible to separate two approaches to multicriteria assessment of the efficiency of distributed wireless systems:

- Associated with bringing the set of individual performance indicators to a single integral indicator.
- Methods of the theory of multiple choice and decision-making with a significant number of individual performance indicators, approximately equally important [4].

## 6 "Zombie" Model of Security Breaches Consideration

Security breaches, based on this model, have a clearly separated stage, as it's shown in Fig. 1. The attack model, used by the attacker, can be presented as follows: shallow study (reconnaissance), in-depth study (scanning of communication channels), complete study (mapping), access to the operating system (OS), extension of authority, "Zombie" OS, manipulation of information, removal of traces of a crime, as well as the installation of spyware software, if it's needed.

The system "zombification" passes through malicious code, which is entered into the OS for remote access. After that, a "zombie" OS runs the next attack and adds new workstations to the "zombie" network (the so-called botnet). At the end of the attack, traces of an attacker's presence in the system are deleted.

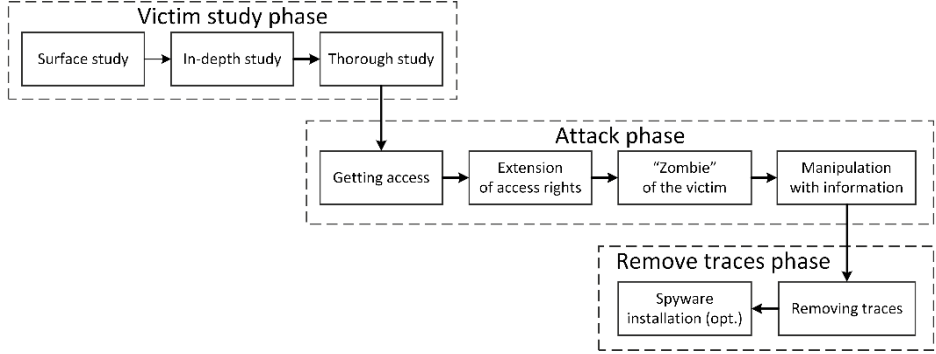

**Fig. 1.** Stages of the "zombie" model.

The "zombie" model efficiency [$c^{-1} \cdot USD^{-1}$] can be calculated by the formula:

$$E = \frac{n \cdot s}{\Delta t \cdot C} \tag{1}$$

where $n$ is the number of potential servers on which the attack is implemented; $s$ is the number of computers that work directly with one server; $\Delta t$ is the time of the system in the "zombie" state; $C$ is the cost of the attack: the cost of writing a botnet, additional costs for the input and distribution of lost code, additional costs [6].

## 7   Formal Attack Model

A formal attack model (AM), within the described above processes, taking into account the proposals [7, 8] can be represented as follows:

$$F_{AM} = \langle M_{AM}^{object}, M_{AM}^{scenario}, M_{AM}^{option} \rangle \tag{2}$$

where $M_{AM}^{object}$ is a component that describes the level of parametrization of the security analysis (SA) process and serves for the establishing the set of analyzed objects, the purposes of the performing attacking actions and the parameters, characterizing the offender. As a rule, it is a pair: object of attack—the purpose of the attack, for example, port scan.

$M_{AM}^{scenario}$ is a component that describes the script level and serves to create a plurality of different scenarios (sequence of attacking actions), taking into account the purpose formed at the level of parameterization of the SA process, which should be achieved by the offender. At the same time, scripting is carried out by the method of a complete overview of all sub-targets of attacking the purpose action, for example, the target "intelligence," sub-targets—"scan of ports," "definition of the OS type," etc.

$M_{AM}^{option}$ is a component that describes all possible variants of the attacker's actions on the basis of its characteristics, also includes an algorithm for the formation of the attack tree.

It, in turn, can be represented as follows:

$$M_{AM}^{option} = F(A, E, F^{option}) \quad (3)$$

where $A$ is set of all attacking actions; $E$ is set of all exploits; $F^{option}$ is set of the functions of this component.

At the same time, the filling of the sets $A$ and $E$ is based on open vulnerability databases, for example, the *Open Source Vulnerability Database or the National Vulnerability Database* (NVD, attacking actions of the implementation stages, enhancing privileges, and implementing the threat), as well as expert knowledge (attacking actions of the stages of intelligence, concealment traces, creation of secret moves).

## 8 Methods for Threats Assessing

### 8.1 Denial of Service Probability

The probability of denial of service data (natural disaster, force majeure, total or partial loss of data, unauthorized access, etc.) allows you to obtain results in the form of a scale of assessments of potential threats and their consequences. The method operates with a set of indicators and for each individual case will be different. The values of the indicators are approximate, based on available statistics or expert estimates, which makes it impossible to analyze it with a small amount of accumulated statistical data [4].

### 8.2 Expected Vulnerability Damage from Threats

Expected vulnerability damage from the $i^{th}$ threat—an empirical method of evaluation, was first proposed by the specialists of IBM [9]:

$$R_i = 10^{S_i + V_i - 4} \quad (4)$$

where $S_i$ and $V_i$ is coefficients that characterize the possible frequency of occurrence of the threat and the value of the possible damage when it occurs (the value of both coefficients—integers in the interval [0, 7], for $S_i$ "0"—almost never, "7"—more than 1,000 times per year; $V_i$ from 1 to 10 million dollars) [4, 6, 9, 10].

This methodology can be described by a system of equations, resulting in parameters at intervals:

$$\begin{cases} R_i = 10^{S_i + V_i - 4} \\ S_i = 7 \cdot 10^{-3} \cdot s_i, \ 0 \leq s_i \leq 10^3 \\ S_i = 7, \ s_i > 10^3 \\ V_i = 7 \cdot 10^{-7} \cdot (v_i - 1), \ 1 \leq v_i \leq 10^7 \\ V_i = 7, \ v_i > 10^7 \end{cases} \quad (5)$$

where $s_i$ is predictable or actual number of attacks per year, $v_i$ is amount of predictable or real damage in monetary units.

In this case, the increase in the second interval is not taken into account; we propose to correct the formula, taking into account the growth in the whole area of determination of characteristics. It is proposed to use a hyperbolic tangent (more precisely, only its positive part in the first quadrant) in the new formula. Based on the characteristics of the hyperbolic tangent function, additional coefficients are introduced:

$$f(x) = k_{max} \cdot \tanh \frac{2x}{b_{max}} \qquad (6)$$

where $k_{max}$ corresponds to the maximum of the scale that is 7, and $b_{max}$ is maximum value of the predictable or real value, the coefficient 2 is introduced for a better scaling by abscissa.

Then the system can be written as follows:

$$\begin{cases} R_i = 10^{S_i+V_i-4} \\ S_i = 7 \cdot \tanh \frac{s_i}{500}, \ 0 \leq s_i \\ V_i = 7 \cdot \tanh \frac{v_i-1}{5 \cdot 10^5}, \ 1 \leq v_i \end{cases} \qquad (7)$$

The formula of expected damage from $i^{th}$ threat can be written in general terms:

$$R_i = 10^{7 \cdot \tanh \frac{s_i}{500} + 7 \cdot \tanh \frac{v_i-1}{5 \cdot 10^5} - 4}, 0 \leq s_i, 1 \leq v_i \qquad (8)$$

This method, as can be seen from the graph (Fig. 2), does not allow to compare different information systems (due to the significant variation in the cost of systems, their scale and workload), since the estimated damage is relative. The method shows the most adequate results in the case of comparing the security of the same system at different points in time or when the state changes its quality.

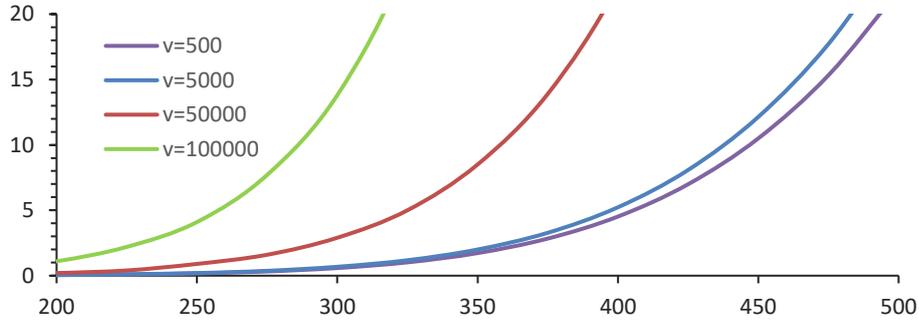

**Fig. 2.** Expected damage quantification.

### 8.3 Set of Values for Defining Security Requirements

The set of values for defining security requirements is another proposed method in [10], which operates with a normalized level of security in the continuum of values [0, 1], and reliability indicators are a function of belonging $\mu^A(x_i)$, where $x_i$ is an element of the *X* set (security requirements), and *A* is a plural of values, defining the fulfillment of security requirements:

$$A = \sum_{i=1}^{N} \frac{\mu^A(x_i)}{x_i} \qquad (9)$$

where $\frac{\mu^A(x_i)}{x_i}$ is normalized pair "function of accessory/element." Then it is possible to evaluate the effectiveness of clearly defined safety criteria.

This method has the main drawback: the system may be evaluated only with a predetermined set of criteria.

### 8.4 Assessing the Threats and Losses

The analytical method for assessing the threats and losses, associated with them, operates with the average indicator of the appearance of the threat *L* and the magnitude of the probability distribution *f(L)*. To estimate the losses, the value *m* with mean deviation *v* is used.

For analysis, it is imperative to have statistics of security breaches and measured values of losses for these attacks.

The issue with the method is the inability to calculate the impact of information security (IS) on *L* and, accordingly, on *m*, and therefore to assess the effectiveness of the measures of IS [4].

### 8.5 Degree of Security

The degree of security provides a rough estimate of the effectiveness of the IS system. The method operates with the subjective coefficients of weight $i^{th}$ characteristic $W_i$ and the ball values of each characteristic $G_i$, which is determined by expert's estimates.

The formula for the degree of security is as follows:

$$S = \frac{1}{N}\sum_{i=1}^{N} W_i \cdot G_i \qquad (10)$$

where *N* is amount of the characteristics.

The method has two drawbacks: it is impossible to compare systems with different sets of characteristics and it does not take into account the dependence of the weighting factor and the value of the characteristic of the characteristic itself [4, 10].

### 8.6 Comprehensive Method of the Information Objects Security

The author of the paper proposes to use normalized characteristic *S\** to assess the degree of the system security, and at the same time, to consider the subjective factors of the importance of the $i^{th}$ characteristic and the ball value of each characteristic, as a function of the characteristics:

$$\begin{cases} W_i = f_W(x_i) \\ G_i = f_G(x_i) \end{cases} \quad (11)$$

where $f_W$ and $f_G$ are functions of the characteristic $x_i$.

The general formula for monotonous $f_W$ and uncertain function $f_G$ is as follow:

$$S^* = \frac{1}{N}\sum_{i=1}^{N} W^*(x_i) \cdot G^*(x_i) \quad (12)$$

where $W^*(x_i)$ is normalized weighting factor of subjective estimation from $x_i$:

$$W^*(x_i) = \left|\frac{f_W(x_i)}{\max[f_W(x_i)]}\right| \quad (13)$$

and $G^*(x_i)$ is normalized score value of a function:

$$G^*(x_i) = \left|\frac{G_i^\Sigma}{G_{i\,max}^\Sigma}\right| \quad (14)$$

Intermediate values of which are defined as integral characteristics:

$$\begin{cases} G_i^\Sigma = \int_{x_i^{begin}}^{x_i^{end}} f_G(x)dx \\ G_{i\,max}^\Sigma = \int_{x_i^{min}}^{x_i^{max}} f_G(x)dx \end{cases} \quad (15)$$

where $x_i^{begin}$ and $x_i^{end}$ are the beginning and the end of the range of values for a given characteristic that exists and is continuous in the range from $x_i^{min}$ to $x_i^{max}$.

In the given case the normalized level of safety of the system will always be $S^* \leq 1$. $S^*$ is "absolutely" protected system, when all the existing characteristics $x_i$ are considered. In the general case, the proposed modification of the method allows to obtain a normalized level of security for any system with a number of characteristics (but not less than 3), and to conduct a comparative analysis of IS in systems with a different set of characteristics.

Because the method operates with the results, obtained through expert evaluation, before the data processing begins, it is necessary to assess the adequacy of the expert group. To assess the adequacy it's needed to determine the coefficient of concordance, which involves the elements of functional-cost analysis.

## 9 Functional-Cost Analysis

Let's suppose we have $N$ essential characteristics that are included in the $X$ set of all characteristics of the system $[x_1, x_2 \ldots x_N] \in X$.

We determine experimentally or analytically the intervals of values for all characteristics (minimum and maximum values), as well as the average value (which does not necessarily coincide with the arithmetic mean and maximum values). In the found intervals, experts determine the point values of each characteristic $G_i$:

$$\begin{cases} G_1 = f_G(x), x = \overline{x_1^{min}, x_1^{av}, x_1^{max}} \\ G_2 = f_G(x), x = \overline{x_2^{min}, x_2^{av}, x_2^{max}} \\ ... \\ G_N = f_G(x), x = \overline{x_N^{min}, x_N^{av}, x_N^{max}} \end{cases} \quad (16)$$

Based on the obtained data, for the sake of clarity, the charts (11), used by experts to determine the following characteristics, are constructed.

### 9.1 Parameters Weighting

The weighting of the parameters is determined by the method of prioritization, according to which the priorities of the characteristics are determined by the expert group ($M$ is the number of experts), and as a the result, the comparison table is compiled (see Table 1), in which the average score is reduced to a numerical form according to the principle: ">" corresponds to 1.5, "="—to 1.0, and "<"—to 0.5.

**Table 1.** Expert evaluation of the parameters importance

| Parameter pairs | Experts | | | | | Average rating | Numeric value |
|---|---|---|---|---|---|---|---|
| | 1 | 2 | 3 | ... | M | | |
| $x_1$ and $x_2$ | = | = | > | ... | > | > | 1.5 |
| $x_1$ and $x_i$ | > | > | > | ... | > | > | 1.5 |
| ... | ... | ... | ... | ... | ... | ... | ... |
| $x_1$ and $x_N$ | > | > | > | ... | > | > | 1.5 |
| $x_2$ and $x_i$ | > | > | < | ... | = | < | 0.5 |
| ... | ... | ... | ... | ... | ... | ... | ... |
| $x_{i-1}$ and $x_i$ | > | < | > | ... | > | > | 1.5 |
| ... | ... | ... | ... | ... | ... | ... | ... |
| $x_{N-1}$ and $x_N$ | > | > | > | ... | > | > | 1.5 |

Due to the received data, a table of the characteristics of the priorities is filled out (see Table 2), in which the coefficient 1.0 is taken for pairs $x_i/x_i$.

**Table 2.** Determination of the characteristics of the priorities

| | Characteristics | | | | | Importance | | Validity | |
|---|---|---|---|---|---|---|---|---|---|
| | $x_1$ | $x_2$ | ... | $x_i$ | ... | $x_N$ | $b_i$ | $\varphi_i$ | $b'_i$ | $W_i = \varphi'_i$ |
| $x_1$ | 1.0 | 1.5 | ... | 1.5 | ... | 1.5 | 7.0 | 0.28 | 34.0 | 0.292 |
| $x_2$ | 0.5 | 1.0 | ... | 1.5 | ... | 1.5 | 5.0 | 0.20 | 22.5 | 0.193 |
| ... | ... | ... | ... | ... | ... | ... | ... | ... | ... | ... |
| $x_i$ | 0.5 | 0.5 | ... | 1.0 | ... | 1.5 | 4.5 | 0.18 | 20.5 | 0.176 |
| ... | ... | ... | ... | ... | ... | ... | ... | ... | ... | ... |
| $x_N$ | 0.5 | 0.5 | ... | 0.5 | ... | 1.0 | 3.0 | 0.12 | 14.0 | 0.120 |
| $\Sigma$ | | | | | | | | 1.0 | | 1.0 |

Degree of importance $\varphi_i$ of each parameter:

$$\varphi_i = \frac{b_i}{\sum_{i=1}^{N} b_i}, \tag{17}$$

$$b_i = \sum_{j=1}^{N} a_{ij} \tag{18}$$

where $b_i$ is the weight of the $i^{th}$ parameter on the basis of expert assessments; $a_{ij}$ is the numerical value of the priority.

The coefficient $W_i$ of the importance of the $i^{th}$ parameter is determined in the second step:

$$W_i = \dot{\varphi}_i = \frac{\dot{b}_i}{\sum_{i=1}^{N} \dot{b}_i}, \tag{19}$$

$$\dot{b}_i = \sum_{j=1}^{N} a_{ij} \cdot b_j \tag{20}$$

### 9.2 Assessment of the Expert Group Adequacy

Assessment of the adequacy of the expert group is carried out after determining the dependence of the ball values of each characteristic of the characteristic itself; the discrete function is reduced to a continuous one from (11).

The sum of the ranks of each parameter:

$$R_i = \sum_{j=1}^{M} r_{ij} \tag{21}$$

where $r_{ij}$ is the rank of the $i^{th}$ characteristics, determined by the $j^{th}$ expert.

Checking the total amount of the ranks, this must be equal:

$$R_{ij} = \frac{1}{2} \cdot M \cdot N \cdot (N+1) \tag{22}$$

The average amount of ranks:

$$R_{av} = \frac{1}{N} \cdot R_{ij} \tag{23}$$

Rejection of the sum of the ranks for each $i^{th}$ characteristic from the average amount (the sum of deviations for all characteristics should be zero):

$$\Delta_i = R_i - R_{av} \tag{24}$$

Total amount of squares of deviations:

$$S = \sum_{i=1}^{N} \Delta_i^2 \tag{25}$$

The coefficient of concordance:

$$W = \frac{12 \cdot S}{M^2 \cdot (N^3 - N)} \tag{26}$$

The coefficient of concordance can take the value $0 \leq W \leq 1$. In the case of complete consistency of expert opinions, the coefficient is $W = 1$. If $W \geq W_{nom}$, the certain data are trustworthy and are usable. For the means of computer technology adopted $W_{nom} = 0.67$, the same value can be used for distributed wireless systems [11, 12]. Since in this case not only wireless systems can be used, the tolerance of the deviation of the values of the concordance coefficient will be taken at the level of ⅕ from its normal value:

$$W_{\text{distribution of communication systems}} = 0.67 \pm 20\% \quad (27)$$

The results of field experiments (Fig. 3) suggest that the recommended number of estimated safety parameters and the number of experts evaluating these parameters are interdependent. This is confirmed by the family of curves constructed on the basis of formula (26).

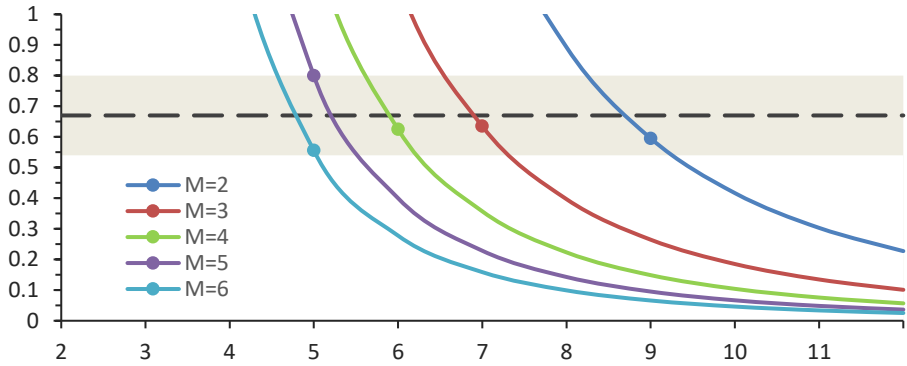

**Fig. 3.** Relationship of security parameters and experts.

## 10   Conclusions and Future Work

The existing models and methods for assessing security and risks, with their drawbacks were considered in the paper. The proposed modifications are intended to improve existing methods and include more precise approximation (for expected damage to the vulnerability) and generalization of the function (for the degree of security). In addition, it is proposed to use elements of functional-cost analysis to verify the reliability of expert evaluation.

From the above, we can say that our method of evaluation is not yet sufficiently thorough and requires more detailed consideration and the introduction of step-by-step instructions in the comprehensive assessment of the security and risks for distributed wireless systems.

The paper describes the sequence of defining the system's security. In the future, we plan to compare the calculation of efficiency and risk.